\documentclass[onecolumn,a4paper,12pt,oneside]{article} 
\usepackage[top=2.5cm,left=2.5cm,right=2.5cm,bottom=3cm]{geometry}    
\usepackage[nostamp]{draftwatermark} % watermark programmed but not printed
\SetWatermarkText{\bf DRAFT}
\SetWatermarkAngle{45}
\SetWatermarkScale{4}
\SetWatermarkLightness{0.85}     
\usepackage[parfill]{parskip}    % Activate to begin paragraphs with an empty line rather than an indent
\usepackage{graphicx}
\usepackage{amssymb}
\usepackage{amsmath}
\usepackage{amsfonts}
\usepackage{amsthm}
\usepackage{epstopdf}
\usepackage[usenames,dvipsnames,svgnames,table]{xcolor}
\usepackage{tikz}
\usepackage{array}
\usepackage[T1]{fontenc}
\usepackage[utf8]{inputenc}
\usepackage{booktabs} % for much better looking tables
\usepackage{array} % for better arrays (eg matrices) in maths
\usepackage{paralist} % very flexible & customisable lists (eg. enumerate/itemize, etc.)
\usepackage{listings} 
\usepackage{verbatim} % adds environment for commenting out blocks of text & for better verbatim
\usepackage{subfig} % make it possible to include more than one captioned figure/table in a single float
\usepackage[colorlinks=true]{hyperref}
\hypersetup{colorlinks=true, linktocpage=true, linkcolor=Blue, citecolor=Green,
hypertexnames=true, urlcolor=MidnightBlue,pdftex}
\usepackage[english]{babel}
\usepackage{wrapfig}
\usepackage{multirow}
\usepackage{quoting}
\usepackage{rotating}
\quotingsetup{font=normalsize}
\theoremstyle{plain}

\usepackage{bm}
\usepackage{abstract}
\usepackage{cite}
\usepackage{float}
\usepackage{textcomp} 	
\usepackage{tabularx} % Better tables
\usepackage{caption}
\usepackage{authblk} % for authors, affiliations etc.
\usepackage{eso-pic} % for background figures
\usepackage{lineno} % for numbering lines
\providecommand{\keywords}[1]{\textbf{{Key words: }} #1} % command for keywords
\newcommand{\be}{\begin{equation}}
\newcommand{\ee}{\end{equation}}
\newcommand{\bsp}{\begin{split}}
\newcommand{\esp}{\end{split}}

\newcommand{\beps}{\boldsymbol{\varepsilon}}
\newcommand{\bk}{\boldsymbol{\kappa}}
\renewcommand{\Phi}{\varPhi}
\newcommand{\gr}{\mathbf}
\newcommand{\eps}{\varepsilon}

\renewcommand{\Theta}{\varTheta}
\renewcommand{\Psi}{\varPsi}
\renewcommand{\Sigma}{\varSigma}
\newcommand{\A}{\mathbb{A}}
\newcommand{\B}{\mathbb{B}}

\newcommand{\D}{\mathbb{D}}

\renewcommand{\Delta}{\varDelta}
\renewcommand{\phi}{\varphi}
\renewcommand{\psi}{\varPsi}

\newcommand{\N}{\mathbf{N}}
\newcommand{\M}{\mathbf{M}}
\newcommand{\te}{t_\eps}
\newcommand{\re}{r_\eps}
\newcommand{\fe}{\phi_\eps}
\newcommand{\tk}{t_\kappa}
\newcommand{\rk}{r_\kappa}
\newcommand{\fk}{\phi_\kappa}

\title{{\LARGE{\bf Elastic bounds of the coupling tensor of anisotropic composite laminates: a polar approach}}\bigskip\\{\small - PREPRINT-\medskip\\ Final, extended version published in Proceedings of the Royal Society A \href{https://dx.doi.org/10.1098/rspa.2024.0050}{https://dx.doi.org/10.1098/rspa.2024.0050}}}
\author{Paolo VANNUCCI\bigskip\\
{\small{\it LMV - Laboratoire de Mathématiques de Versailles UMR8100\\
 UVSQ - Université de Versailles et Saint Quentin\\
           45, Avenue des Etats-Unis, 78035 Versailles - France}\\
         \href{mailto:paolo.vannucci@uvsq.fr}{paolo.vannucci@uvsq.fr}}}
\begin{document}
\maketitle

%%%%%%%%%%%% ABSTRACT %%%%%%%%%%%%%%%%%
\hrule
\begin{abstract}
In the equivalent single layer theories of anisotropic laminates, the coupling tensor $\B$ describes the relation between the in- and out-of plane behavior of the plate. This tensor has some peculiar characteristics and in particular it is not defined, so it is normally considered as an unbounded tensor. We show in this paper a way to determine some relations between the polar invariants of $\B$ and those of the tensors describing the in- and -out-of plane behaviors of the laminate, $\A$ and $\D$ respectively. These relations constitute a set of bounds for the invariants of all the tensors of the laminate, $\A,\B$ and $\D$. It is hence shown that also the components of $\B$ must satisfy some conditions.  Some peculiar cases, interesting for applications, are also considered. \\

% keywords
\end{abstract}
\medskip
\hrule
\bigskip

%%%%%%%%%%% TEXT of The ARTICLE %%%%%%%%%%%%

\section{Introduction}
\label{sec:intro}
%The problem of determining whether or not the moduli of the coupling tensor $\B$ of an anisotropic laminate is addressed in this paper. 

In the classical theory of elastic laminates, the  behavior of the plate is described by a law of the type
\be
 \label{eq:fundlaw}
 \left\{\begin{array}{c}\gr{N} \\ \gr{M}\end{array}\right\}=
 \left[\begin{array}{cc}h\A & \dfrac{h^2}{2}\B\smallskip\\ \dfrac{h^2}{2}\B & \dfrac{h^3}{12}\D\end{array}\right]
 \left\{\begin{array}{c}\beps \\ \bk\end{array}\right\},
\ee
where, \cite{jones,vannucci_libro}, $h$ is the plate's thickness, $\N,\M$ are respectively the tensor of membrane forces and bending moments, $\beps,\bk$ the extension and curvature tensors, $\A,\D$ the stiffness tensors of the extension and bending behaviors and $\B$ the coupling tensor. $\A,\B,\D$ are tensors of the elastic type, i.e. fourth-rank tensors with the major and minor symmetries.  In other  equivalent single layer models for laminated plates, namely in the theory of Reissner-Mindlin, \cite{Reissner45,Mindlin51} or in the third-order shear deformation theory of Reddy, \cite{Reddy84,Reddy87,Reddy99}, the definition of the curvatures changes and other tensors, describing the transverse shear behavior, are to be included, but the definition, and the properties, of $\A,\B$ and $\D$ do not change.

A deep investigation of the mechanical and mathematical properties of $\B$ has been presented in \cite{vannucci23a}. Moreover, in \cite{vannucci12joe1,vannucci23b,vannucci23c}, also the thermal coupling effects have been investigated.

Unlike $\A$ and $\D$, that are positive definite, $\B$ is not defined. At least, this is what is commonly admitted, and not only for $\B$, but also for other coupling tensors describing some sort of coupling between two physical phenomena, e.g. in the theory of quasi-crystals, \cite{mariano}. Actually, the positive definiteness of $\A$ and $\D$, like that of any other elastic tensor, is a consequence of the  work, necessarily positive, done by the external forces. In the case of $\A$ and $\D$, this is computed considering the laminate as uncoupled. The case of coupled laminates, i.e. having $\B\neq\mathbb{O}$, is almost unconsidered in the literature and $\B$ is commonly seen as an {\it undefined} tensor, i.e. in general not positively nor negatively defined. 

The positive definiteness of an elastic tensor gives as a result some bounds on the elastic moduli of the tensor in a given mathematical representation, e.g. the well known bounds on the Lamé's constants or on the Young's modulus and the Poisson's ratio for isotropic materials. More delicate is the case of anisotropic materials, \cite{jones,ting}, for which a clear and definite set of bounds for the elastic moduli has not yet been defined in  three-dimensional elasticity for any possible elastic syngony, namely for the most general one of triclinic materials. However, the same problem has been solved definitely in two-dimensional elasticity, using the polar formalism, \cite{meccanica05,vannucci14mmas,vannucci_libro}. By this mathematical technique, introduced as early as 1979 by G. Verchery, \cite{verchery79}, any elastic tensor is represented by invariant moduli and angles. In this way, the bounds on the elastic tensor are given on its invariants, so intrinsically representing the elastic limits of the material. 
The same procedure has been used to determine the so-called {\it geometrical bounds}, \cite{vannucci13}, i.e. the bounds determining the elastic domain for a laminate when the stacking is considered. 

The question of the bounds on $\B$ remains open. This paper addresses exactly this subject and in particular it proposes an approach to give a first answer  the following questions:
\begin{enumerate}[i.]
\item is it possible to establish some bounds for the moduli of $\B$?
\item how the known bounds on $\A$ and $\D$ are modified when $\B\neq\mathbb{O}$?
\item is it possible to give an explicit form to the bounds on the moduli of $\B$?
\item can such bounds be given, in all the cases, in an invariant form?
\item how the existence of some peculiar circumstances, e.g. a material symmetry, does affect the bounds on $\B$?
\end{enumerate}

This research has two motivations: the first one, is a purely scientific question, interesting {\it per se}, to fill a gap still existing in the scientific literature: is it possible to give some bounds to tensor $\B$? Another motivation  can be found in optimization problems: any design problem  for a coupled laminate is correctly formulated only if the design space is well determined, so as to  properly define the feasibility domain. 

The paper is organized as follows: in the next Section, the polar formalism is briefly recalled, especially for representing tensors $\A,\B,\D$. Then, the  procedure used to determine the bounds on $\B$ is detailed and subsequently the attempt to establish a general solution is explained. Some special cases are then considered and finally some conclusion is drawn in the end. 

\section{Recall of the polar formalism for a laminate}
\begin{figure}
\includegraphics[width=\textwidth]{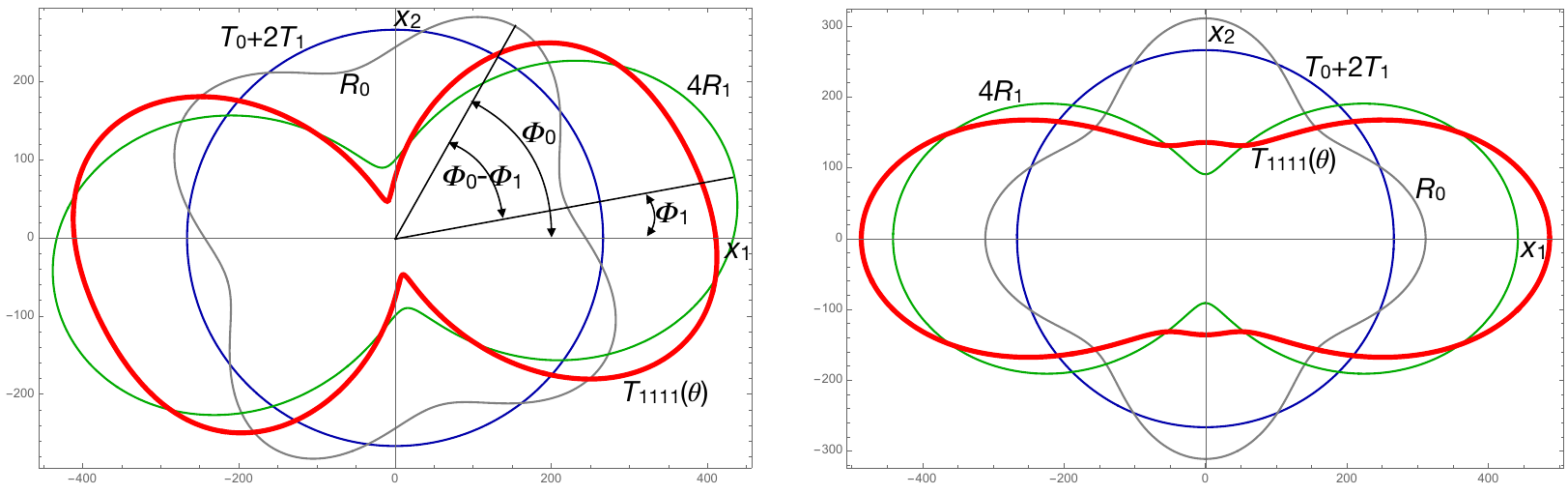}
\caption{The decomposition of anisotropic plane elasticity in the polar method for a glass-epoxy layer with $T_0=92.38$ MPa, $T_1=86.97$ MPa, $R_0=44.86$ MPa, $R_1=43.82$ MPa, source \cite{MILHDBK}; left: a completely anisotropic material obtained if it was $\Phi_0=\pi/3, \Phi_1=\pi/20$, with, in blue (indicated also by $T_0+2T_1$), the isotropic phase, in gray (indicated by $R_0$) the $R_0$  phase, in green (indicated by $4R_1$) the $R_1$ phase and in thick red the overall result, i.e. the component $T_{1111}(\theta)$. Right:   the true, orthotropic material, corresponding to $\Phi_0=\Phi_1=0$.}
\label{fig:1}
\end{figure}

The reader is referred to \cite{vannucci_libro,meccanica05} for a complete presentation of the polar formalism. Here, only major results of this theory, relevant to the topic of this paper, are presented, without proofs. 
For a given plane elastic tensor $\mathbb{T}$, the polar formalism allows to express the cartesian components at a direction $\theta$ as
\begin{equation}
\label{eq:mohr4}
{\begin{split}
&{T_{1111}{(}\theta{)}{=}{T}_{0}{+}{2}{T}_{1}{+}{R}_{0}\cos{4}\left({{{\varPhi}}_{0}{-}\theta}\right){+}{4}{R}_{1}\cos{2}\left({{{\varPhi}}_{1}{-}\theta}\right)},\\
&{T_{1112}{(}\theta{)}{=}{R}_{0}\sin{4}\left({{{\varPhi}}_{0}{-}\theta}\right){+}{2}{R}_{1}\sin{2}\left({{{\varPhi}}_{1}{-}\theta}\right)},\\
&{T_{1122}{(}\theta{)}{=}{-}{T}_{0}{+}{2}{T}_{1}{-}{R}_{0}\cos{4}\left({{{\varPhi}}_{0}{-}\theta}\right)},\\
&{T_{1212}{(}\theta{)}{=}{T}_{0}{-}{R}_{0}\cos{4}\left({{{\varPhi}}_{0}{-}\theta}\right)},\\
&{T_{1222}{(}\theta{)}{=}{-}{R}_{0}\sin{4}\left({{{\varPhi}}_{0}{-}\theta}\right){+}{2}{R}_{1}\sin{2}\left({{{\varPhi}}_{1}{-}\theta}\right)},\\
&{T_{2222}{(}\theta{)}{=}{T}_{0}{+}{2}{T}_{1}{+}{R}_{0}\cos{4}\left({{{\varPhi}}_{0}{-}\theta}\right){-}{4}{R}_{1}\cos{2}\left({{{\varPhi}}_{1}{-}\theta}\right)}.
\end{split}}
\end{equation}
What is important to remark is the fact that the moduli $T_0,T_1,R_0,R_1$ as well as the difference of the angles $\Phi_0-\Phi_1$ are tensor invariants; the value of one of the two polar angles, usually $\Phi_1$, fixes the frame. To be remarked that the polar method allows for a decomposition of anisotropic 2D elasticity into different {\it elastic phases}: an {\it isotropic phase}, characterized by the two invariants $T_0$ and $T_1$, and two {\it anisotropic phases}, whose amplitudes are determined by the invariants $R_0$ and $R_1$; these two anisotropy phases are shifted of the angle $\Phi_0-\Phi_1$, the fifth tensor invariant. It is worth noting that any rotation of the angle $\theta$ is simply done in the polar method: it is sufficient to subtract $\theta$ from each one of the two polar angles. A general sketch of the decomposition of elasticity into elastic phases is given in Fig. \ref{fig:1}, where the angles $\Phi_0,\Phi_1$ and $\Phi_0-\Phi_1$ are also indicated.

\begin{figure}
\includegraphics[width=\textwidth]{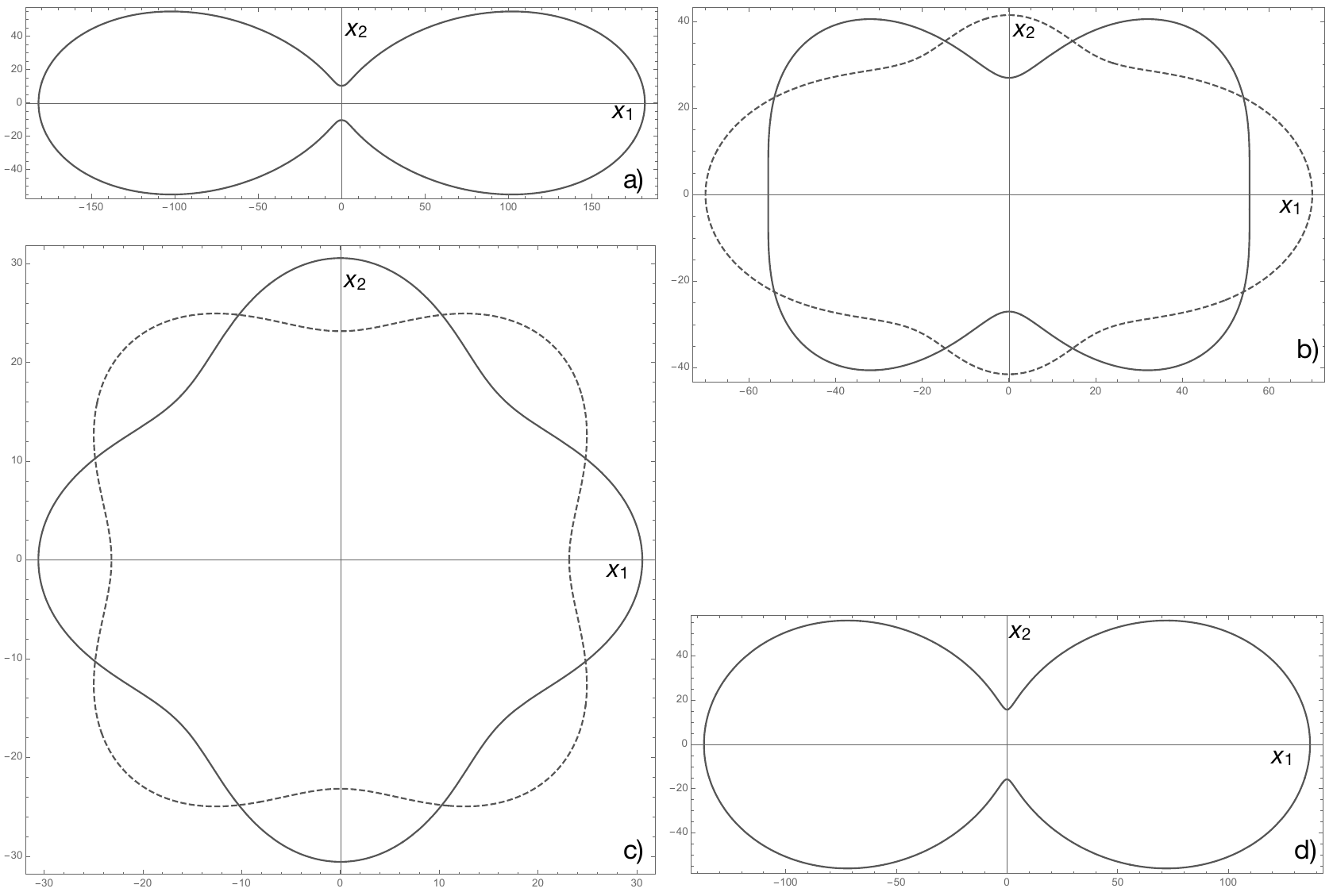}
\caption{Polar diagrams of $T_{1111}(\theta)$ for some examples of elastic symmetries in 2D elasticity; a) a $k=0$ orthotropic ply (T300/5208 carbon epoxy ply, \cite{TsaiHahn} ); b) a $k=1$ orthotropic ply (braided carbon-epoxy BR45a ply, \cite{Falzon98}); with a dashed line: a material with the same moduli but with $k=0$; c) a square symmetric ply, $R_1=0$ (carbon epoxy balanced fabric, \cite{gay14}); with a dashed line: the same material but with $\Phi_0=\pi/4$; d) a $R_0$-orthotropic material, obtained superposing two T300/5208 carbon epoxy plies rotated of $\pi/4$.}
\label{fig:2}
\end{figure}

In the polar formalism, the elastic symmetries are determined by the following values of the invariants:
\begin{enumerate}[i.]
\item ordinary orthotropy: $\Phi_0-\Phi_1=k\dfrac{\pi}{4}, \ k\in\{0,1\}$;
\item $R_0$-orthotropy: $R_0=0$, \cite{vannucci02joe};
\item square symmetry: ${R_1=0}$;
\item isotropy: $R_0=R_1=0$.
\end{enumerate}
These elastic symmetries are the only possible ones in 2D elasticity; in particular, we see that two cases of ordinary orthotropy can exist, sharing the same values of the invariants $T_0,T_1,R_0,R_1$. Moreover,  square symmetry is the 2D corresponding of the 3D cubic syngony: it is the case of a layer having two couples of mutually orthogonal symmetry axes, rotated of $\pi/4$, and with the same values of the elastic moduli along the two orthogonal axes. It is, namely, the case of composite plies reinforced by balanced fabrics, i.e. by fabrics having the same amount of fibers in warp and weft, \cite{vincenti01}. Finally, the case of $R_0$-orthotropy has been discovered for the first time in 2D elasticity thanks to the polar method, \cite{vannucci02joe}, and later also in 3D elasticity, \cite{forte05}. In Fig. \ref{fig:2}, some examples of the above possible cases are given.

The above polar transformations apply also to tensors $\A,\B$ and $\D$; in particular, when a laminate is composed by identical layers, the case considered in this research, \cite{vannucci01ijss, vannucci01joe}, then (we indicate by a superscript $A,B$ or $D$ a polar quantity of $\A,\B$ or $\D$ respectively, while a polar quantity of the basic layer has no superscript)
\be
T_0^A=T_0^D=T_0,\ T_1^A=T_1^D=T_1,\ T_0^B=T_1^B=0.
\ee
The polar formalism can be applied also to other situations, e.g. to the piezoelectric tensor, \cite{vannucci07ijss}, or to thermo-elastic problems, \cite{vannucci07,vannucci12joe1}. Some special cases of elastic non-classical materials have also been studied through the polar formalism, \cite{vannucci10ijss,vannucci10joe,vannucci2016}, however in this research we will consider only classical elastic tensors.

Regarding a second-rank symmetric tensor {\bf L}, in the polar formalism it is
\be
\label{eq:mohr}
\begin{split}
&L_{11}(\theta)=T+R\cos2(\varPhi-\theta),\\
&L_{12}(\theta)=R\sin2(\varPhi-\theta),\\
&L_{22}(\theta)=T-R\cos2(\varPhi-\theta),
\end{split}
\ee
with $T, R$ two invariants and $\Phi$ an angle determined by the choice of the frame. About these relations, we can rephrase the comments given for eq. (\ref{eq:mohr4}), adding that actually they correspond to the analytical expression of the well known graphical construction of the Mohr's circle, \cite{timoshenko1}. In the following, we will indicate by $\te,\re,\fe$ the polar components of $\beps$ and by $\tk,\rk,\fk$ those of $\bk$.

\section{Statement of the problem}
For a coupled laminate, i.e. with $\B\neq\mathbb{O}$, the density of the elastic energy per unit of area of the plate is
\be
U=\frac{1}{2}\left\{\begin{array}{c}\textbf{N}\\\textbf{M}\end{array}\right\}\cdot\left\{\begin{array}{c}\beps\\\bk\end{array}\right\}=\frac{1}{2}\left\{\begin{array}{c}\beps\\\bk\end{array}\right\}\cdot
 \left[\begin{array}{cc}h\A & \dfrac{h^2}{2}\B\smallskip\\ \dfrac{h^2}{2}\B & \dfrac{h^3}{12}\D\end{array}\right]
\left\{\begin{array}{c}\beps\\\bk\end{array}\right\},
\ee
i.e., for being $\B=\B^\top$ by virtue of the major symmetries, \cite{vannucci_alg}, 
\be
U=\frac{h}{24}(12\ \beps\cdot\A\beps+12h\ \beps\cdot\B\bk+h^2\bk\cdot\D\bk).
\ee
The energy density $U$ must be positive for each possible strain state, i.e. $\forall\beps,\bk$; this is the condition leading to express the bounds for the elastic moduli. Following an approach first introduced by Verchery and detailed in  \cite{vannucci_libro,vannucci15ijss}, we express all the tensors in the previous equation by their polar components, once fixed $\theta=0$. Some standard passages lead to
\be
\begin{split}
U&=2h\{2T_1\te^2+[T_0+R_0^A\cos4(\Phi_0^A-\fe)]\re^2+4R_1^A\te\re\cos2(\Phi_1^A-\fe)\}\\
&+2h^2\{2R_1^B\te\rk\cos2(\Phi_1^B-\fk)+2R_1^B\tk\re\cos2(\Phi_1^B-\fe)\\
&+R_0^B\re\rk\cos2(2\Phi_0^B-\fe-\fk)\}\\
&+\frac{h^3}{6}\{2T_1\tk^2+[T_0+R_0^D\cos4(\Phi_0^D-\fk)]\rk^2+4R_1^D\tk\rk\cos2(\Phi_1^D-\fk)\}.
\end{split}
\ee
In this expression the first term in curly braces is due to extension, the second one to coupling and the third one to bending. Let us order the polar moduli of $\beps$ and $\bk$ as a column vector $\{v\}$:
\be
\label{eq:v}
\{v\}=\left\{\begin{array}{c}\te\\\re\\\tk\\\rk\end{array}\right\};
\ee
then, we can rewrite $U$ as the quadratic form
\be
U=\frac{h}{24}\{v\}^\top[M]\{v\},
\ee
where $[M]$ is the $4\times4$ symmetric matrix
\be
\label{eq:matrixM}
\hspace{-25mm}
[M]=
\begin{footnotesize}
\left[
\begin{array}{cccc}
96T_1&96R_1^A\cos2(\Phi_1^A-\fe)&0&48hR_1^B\cos2(\Phi_1^B-\fk)\\
96R_1^A\cos2(\Phi_1^A-\fe)&48[T_0+R_0^A\cos4(\Phi_0^A-\fe)]&48hR_1^B\cos2(\Phi_1^B-\fe)&24hR_0^B\cos2(2\Phi_0^B-\fe-\fk)\\
0&48hR_1^B\cos2(\Phi_1^B-\fe)&8h^2T_1&8h^2R_1^D\cos2(\Phi_1^D-\fk)\\
48hR_1^B\cos2(\Phi_1^B-\fk)&24hR_0^B\cos2(2\Phi_0^B-\fe-\fk)&8h^2R_1^D\cos2(\Phi_1^D-\fk)&4h^2[T_0+R_0^D\cos4(\Phi_0^D-\fk)]
\end{array}
\right]
\end{footnotesize}.
\ee
So, $U>0\ \forall\beps,\bk\iff[M]$ is positive definite $\forall\fe,\fk$. This happens if and only if, see \cite{hohn58} p. 340, 
\be
\label{eq:M1}
M1=96T_1>0,
\ee
\be
\label{eq:M2}
\hspace{-3mm}M2=\det\begin{small}
\left[
\begin{array}{cc}
96T_1&96R_1^A\cos2(\Phi_1^A-\fe)\\
96R_1^A\cos2(\Phi_1^A-\fe)&48[T_0+R_0^A\cos4(\Phi_0^A-\fe)]
\end{array}
\right]\end{small}>0\ \forall\fe,
\ee
\be
\label{eq:M3}
\hspace{-15mm}M3=\det\begin{small}\left[
\begin{array}{ccc}
96T_1&96R_1^A\cos2(\Phi_1^A-\fe)&0\\
96R_1^A\cos2(\Phi_1^A-\fe)&48[T_0+R_0^A\cos4(\Phi_0^A-\fe)]&48hR_1^B\cos2(\Phi_1^B-\fe)\\
0&48hR_1^B\cos2(\Phi_1^B-\fe)&8h^2T_1
\end{array}
\right]\end{small}>0\ \forall \fe,
\ee
\be
\label{eq:M4}
M4=\det[M]>0\ \forall\fe,\fk.
\ee
The first condition, eq. (\ref{eq:M1}), is redundant: $T_1>0$ is a general condition for this polar modulus, see \cite{vannucci_libro} p. 154. Because $T_1$ is a modulus of the basic layer, i.e. of a real material, this condition is automatically satisfied. The other conditions on $M2,M3$ and $M4$ are discussed in the next Section. Before doing that, we notice that only the condition on $M4$ concerns also the curvature field, while those on $M2$ and $M3$ depend just on the extension field.

\section{General solution}
We consider first the different conditions on $M2, M3$ and $M4$, then we examine them together in the next Section.
\subsection{Condition on M2}
\label{sec:M2}
Equation (\ref{eq:M2}) gives the following condition for the positive definiteness of $[M]$:
\be
\label{eq:condM2_1}
T_1[T_0+R_0^A\cos4(\Phi_0^A-\fe)]-2{R_1^A}^2\cos^22(\Phi_1^A-\fe)>0\ \forall\fe.
\ee

This condition is exactly the same already found in \cite{vannucci15ijss} for proving the bounds on the polar moduli of an anisotropic layer; here, the condition refers to tensor $\A$. Following the same steps outlined in \cite{vannucci_libro} or in \cite{vannucci15ijss}, it can be proved that eq. (\ref{eq:condM2_1}) is equivalent to the three conditions
\be
\begin{split}
&T_0-R_0^A>0,\\
&T_0T_1-{R_1^A}^2>0,\\
&T_1(T_0^2-{R_0^A}^2)-2{R_1^A}^2(T_0-R_0^A\cos4\Phi_A)>0,
\end{split}
\ee
where $\Phi_A=\Phi_0^A-\Phi_1^A$, an invariant of $\A$. Actually, as shown in \cite{vannucci_libro,vannucci15ijss}, the second condition above is less restrictive than the third one, so that it can be discarded. %, remembering that  $R_0$ and $R_1$ of an elastic-type tensor are the modulus of a complex number, and as such non negative quantities, \cite{vannucci_libro}, p. 144, 
Finally, the condition on $M2$ becomes
\be
\label{eq:condM2_2}
\begin{split}
&T_0-R_0^A>0,\\
&T_1(T_0^2-{R_0^A}^2)-2{R_1^A}^2(T_0-R_0^A\cos4\Phi_A)>0.
\end{split}
\ee
We remark that conditions (\ref{eq:condM2_2}) are written in terms of tensor invariants of $\A$, so they are frame independent. 

Of course, the order in which the components of $\beps$ and of $\bk$ appear in vector $\{v\}$, eq. (\ref{eq:v}) is completely arbitrary, any other order being allowed. In particular, if one choses the order putting first the components of $\bk$, then those of $\beps$, i.e. if we had put
\be
\label{eq:v2}
\{v\}=\left\{\begin{array}{c}\tk\\\rk\\\te\\\re\end{array}\right\},
\ee
then we had found for $M2$ a similar expression, but with the index $D$ replacing the index $A$ everywhere, i.e. the conditions would concern in this case the moduli of $\D$ and not those of $\A$. Because of the arbitrariness of the order of the components of $\{v\}$, we can conclude that actually the above conditions (\ref{eq:condM2_1}) and (\ref{eq:condM2_2}) must hold necessarily also for the components of $\D$:
\be
\label{eq:condM2_1bis}
T_1[T_0+R_0^D\cos4(\Phi_0^D-\fk)]-2{R_1^D}^2\cos^22(\Phi_1^D-\fk)>0\ \forall\fk
\ee
and
\be
\label{eq:condM2_3}
\begin{split}
&T_0-R_0^D>0,\\
&T_1(T_0^2-{R_0^D}^2)-2{R_1^D}^2(T_0-R_0^D\cos4\Phi_D)>0,
\end{split}
\ee
with of course $\Phi_D=\Phi_0^D-\Phi_1^D$.

\subsection{Condition on M3}
\begin{figure}
\begin{center}
\includegraphics[width=.7\textwidth]{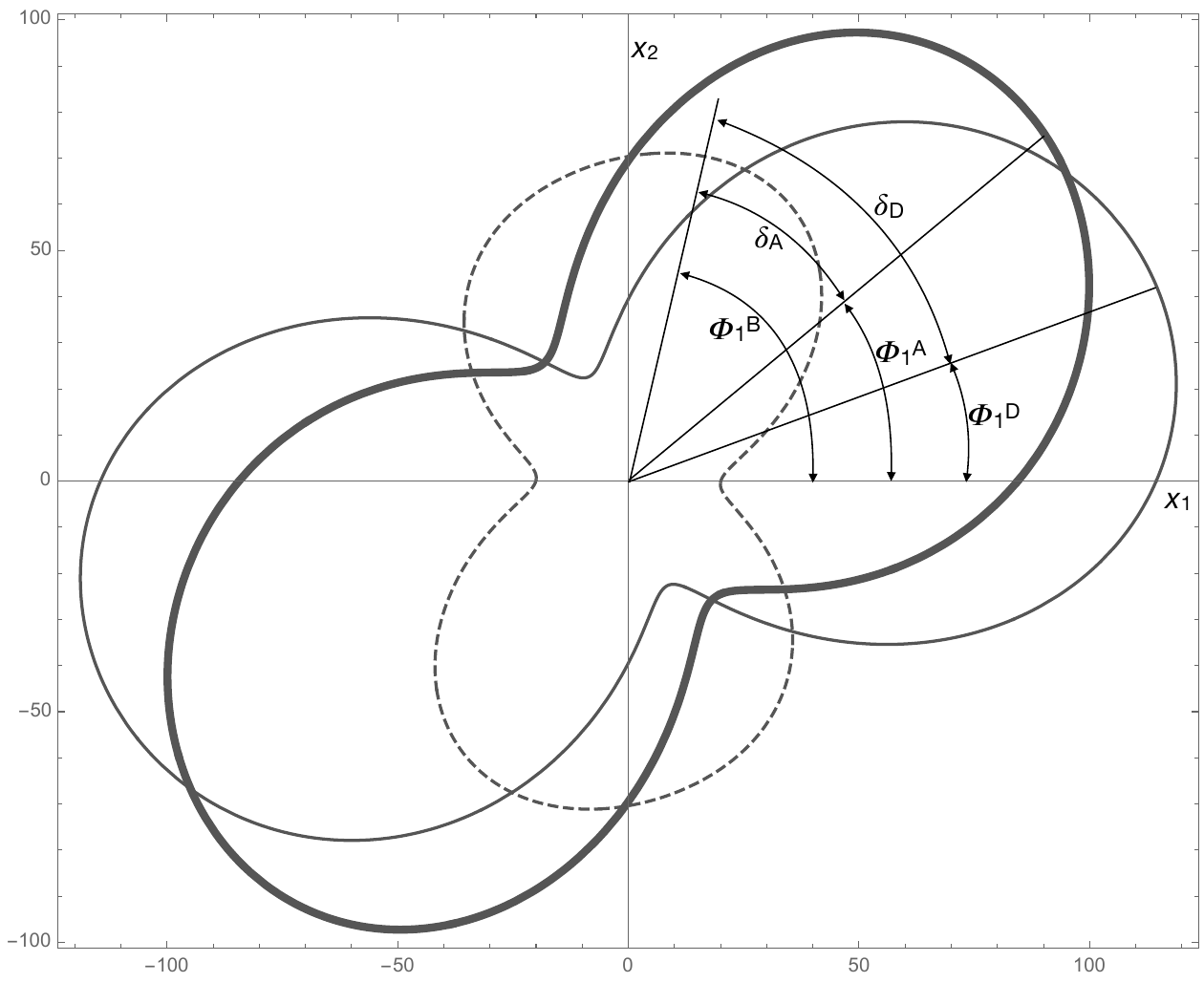}
\caption{Polar diagrams of $A_{1111}(\theta)$, thick line, $B_{1111}(\theta)$, dashed line,  and $D_{1111}(\theta)$, thin line, with indication of the polar  angles $\Phi_1^A,\Phi_1^B,\Phi_1^D$ and of the shift angles $\delta_A$ and $\delta_D$, for the general case of a completely anisotropic coupled laminate.}
\label{fig:3}
\end{center}
\end{figure}

Developing the determinant in eq. (\ref{eq:M3}), we get the condition
\be
\label{eq:condM3_1}
\begin{split}
h^2T_1[T_0T_1+T_1R_0^A\cos4(\Phi_0^A-\fe)-2{R_1^A}^2\cos^22(\Phi_1^A-\fe)&\\-6{R_1^B}^2\cos^22(\Phi_1^B-\fe)]&>0\ \forall\fe.
\end{split}
\ee
To solve this condition, we consider just the term in square brackets, as $T_1>0$, as already said; moreover, we introduce the angles
\be
\alpha=\Phi_1^A-\fe\ \Rightarrow\ \Phi_0^A-\fe=\Phi_A+\alpha,
\ee
and
\be
\delta_A=\Phi_1^B-\Phi_1^A\ \Rightarrow\ \Phi_1^B-\fe=\delta_A+\alpha.
\ee
The angle $\delta_A$ is the {\it shift angle} of tensor $\A$ with respect to tensor $\B$, whose geometrical meaning is shown by Fig. \ref{fig:3}. With this, eq. (\ref{eq:condM3_1}) is equivalent to
\be
T_1[T_0+R_0^A\cos4(\Phi_A+\alpha)]>2{R_1^A}^2\cos^22\alpha+6{R_1^B}^2\cos^22(\delta_A+\alpha)>0\ \forall\alpha.
\ee
Using standard trigonometry, this can be transformed first to
\be
\begin{split}
&T_0T_1-{R_1^A}^2-3{R_1^B}^2>(T_1R_0^A\sin4\Phi_A+3{R_1^B}^2\sin4\delta_A)\sin4\alpha-\\
&-(T_1R_0^A\cos4\Phi_A-3{R_1^B}^2\cos4\delta_A-{R_1^A}^2)\cos4\alpha\ \ \ \forall\alpha,
\end{split}
\ee
then to
\be
\label{eq:condM3_2}
\begin{split}
&T_0T_1-{R_1^A}^2-3{R_1^B}^2>[(T_1R_0^A\sin4\Phi_A+3{R_1^B}^2\sin4\delta_A)^2+\\&+(T_1R_0^A\cos4\Phi_A-3{R_1^B}^2\cos4\delta_A-{R_1^A}^2)^2]^{\frac{1}{2}}\cos4(\alpha+\omega)\ \ \ \forall\alpha,
\end{split}
\ee
with
\be
\omega=\frac{1}{4}\arctan\frac{T_1R_0^A\sin4\Phi_A+3{R_1^B}^2\sin4\delta_A}{T_1R_0^A\cos4\Phi_A-3{R_1^B}^2\cos4\delta_A-{R_1^A}^2}.
\ee
For being  satisfied for each possible value of $\alpha$, eq. (\ref{eq:condM3_2}) gives the condition
\be
\begin{split}
&T_0T_1-{R_1^A}^2-3{R_1^B}^2>[(T_1R_0^A\sin4\Phi_A+3{R_1^B}^2\sin4\delta_A)^2+\\&+(T_1R_0^A\cos4\Phi_A-3{R_1^B}^2\cos4\delta_A-{R_1^A}^2)^2]^{\frac{1}{2}},
\end{split}
\ee
solved into the two conditions
\be
\label{eq:condM3_3}
T_0T_1-{R_1^A}^2-3{R_1^B}^2>0
\ee
and
\be
\label{eq:condM3_3bis}
\begin{split}
&(T_0T_1-{R_1^A}^2-3{R_1^B}^2)^2>(T_1R_0^A\sin4\Phi_A+3{R_1^B}^2\sin4\delta_A)^2+\\&+(T_1R_0^A\cos4\Phi_A-3{R_1^B}^2\cos4\delta_A-{R_1^A}^2)^2.
\end{split}
\ee
This last can be easily transformed to
\be
\label{eq:condM3_4}
\begin{split}
%&T_1[T_1(T_0^2-{R_0^A}^2)-2T_0({R_1^A}^2+3{R_1^B}^2)+2R_0^A({R_1^A}^2\cos4\Phi_A+\\
%&+3{R_1^B}^2\cos4(\Phi_A+\delta_A))]+6{R_1^A}^2{R_1^B}^2(1-\cos4\delta_A)>0.
&T_1^2\left(T_0^2-{R_0^A}^2\right)+6T_1R_0^A{R_1^B}^2\cos4\left(\Phi_A+\delta_A\right)-\\
&-2{R_1^A}^2\left[T_1\left(T_0-R_0^A\cos4\Phi_A\right)-3{R_1^B}^2\right]-\\
&-6{R_1^B}^2\left(T_0T_1+{R_1^A}^2\cos4\delta_A\right)>0.
\end{split}
\ee
Equations (\ref{eq:condM3_3}) and (\ref{eq:condM3_4}) are two bounds into which the condition on $M3$, eq. (\ref{eq:M3}), is reduced. Both of them concern moduli of $\A$ and $\B$, so these bounds describe the interactions existing between the moduli of these two tensors. Unlike eq. (\ref{eq:condM3_3}), depending exclusively upon tensor invariants, the condition in eq. (\ref{eq:condM3_4}) depends also on $\delta_A$, which is not a tensor invariant. However, $\delta_A$ is a shift angle, so as such frame independent. Hence, also these bounds, corresponding to the condition $M3$, are frame independent like those deriving from the condition on $M2$.

Using once more the remark done in the previous Section about the order to give to the components of vector $\{v\}$ and operating like before, two more necessary bounds, concerning the moduli of $\B$ and $\D$ can be written as well, completely similar to eqs. (\ref{eq:condM3_3}) and (\ref{eq:condM3_4}):
\be
\label{eq:condM3D}
\begin{split}
&T_0T_1-{R_1^D}^2-3{R_1^B}^2>0,\\
%&T_1[T_1(T_0^2-{R_0^D}^2)-2T_0({R_1^D}^2+3{R_1^B}^2)+2R_0^D({R_1^D}^2\cos4\Phi_D+\\
%&+3{R_1^B}^2\cos4(\Phi_D+\delta_D))]+6{R_1^D}^2{R_1^B}^2(1-\cos4\delta_D)>0,
&T_1^2\left(T_0^2-{R_0^D}^2\right)+6T_1R_0^D{R_1^B}^2\cos4\left(\Phi_D+\delta_D\right)-\\
&-2{R_1^D}^2\left[T_1\left(T_0-R_0^D\cos4\Phi_D\right)-3{R_1^B}^2\right]-\\
&-6{R_1^B}^2\left(T_0T_1+{R_1^D}^2\cos4\delta_D\right)>0,
\end{split}
\ee
where $\delta_D$ is the {\it shift angle} between tensors $\B$ and $\D$, see again Fig. \ref{fig:3}:
\be
\delta_D=\Phi_1^B-\Phi_1^D.
\ee

\subsection{Condition on M4}
The development of the determinant in eq. (\ref{eq:M4}) leads to a condition much more cumbersome to be solved than the previous ones, for two reasons: on the one hand, the expression of this determinant is very complicate and, on the other hand, it must be positive for each possible value of $\fe$ and of $\fk$, i.e. there are two possible independent fields to be taken into account simultaneously: strains and curvatures.

Developing the determinant in eq. (\ref{eq:M4}) leads to the following condition

\be
\begin{split}
&T_0^2 T_1^2\hspace{-1mm}-\hspace{-1mm}2 {R_1^A}^2 T_0 T_1 \cos^22(\varPhi_1^A\hspace{-1mm}-\hspace{-1mm}\varphi_\varepsilon)+{R_0^D} T_0
T_1^2 \cos4 (\varPhi_0^D\hspace{-1mm}-\hspace{-1mm}\varphi_\kappa )-\hspace{-1mm}\\
&\hspace{-1mm}-\hspace{-1mm}2 {R_0^D} {R_1^A}^2 T_1 \cos^22(\varPhi_1^A\hspace{-1mm}-\hspace{-1mm}\varphi_\varepsilon) \cos4 ({\varPhi_0^D}\hspace{-1mm}-\hspace{-1mm}\varphi_\kappa )-\hspace{-1mm}\\&\hspace{-1mm}-\hspace{-1mm}6 {R_1^B}^2 T_0 T_1 \cos^22 (\varPhi_1^B\hspace{-1mm}-\hspace{-1mm}\varphi_\kappa )
+6 {R_1^B}^2 \cos^22 (\varPhi_1^B\hspace{-1mm}-\hspace{-1mm}\varPhi
\varepsilon ) \times\\
&\times\left[3 {R_1^B}^2\hspace{-1mm}-\hspace{-1mm}T_0 T_1\hspace{-1mm}-\hspace{-1mm}{R_0^D} T_1 \cos4 (\varPhi_0^D\hspace{-1mm}-\hspace{-1mm}\varphi_\kappa )+3 {R_1^B}^2 \cos4
(\varPhi_1^B\hspace{-1mm}-\hspace{-1mm}\varphi_\kappa )\right]\hspace{-1mm}-\hspace{-1mm}\\
&\hspace{-1mm}-\hspace{-1mm}{R_0^A} T_1 \cos4 (\varPhi_0^A\hspace{-1mm}-\hspace{-1mm}\varphi_\varepsilon
) \left[3 {R_1^B}^2+{R_1^D}^2\hspace{-1mm}-\hspace{-1mm}T_0 T_1\hspace{-1mm}-\hspace{-1mm}{R_0^D} T_1 \cos4 (\varPhi_0^D\hspace{-1mm}-\hspace{-1mm}\varphi_\kappa )+\right.\\
&\left.+3 {R_1^B}^2 \cos4(\varPhi_1^B\hspace{-1mm}-\hspace{-1mm}\varphi_\kappa )
+{R_1^D}^2 \cos4 (\varPhi_1^D\hspace{-1mm}-\hspace{-1mm}\varphi_\kappa )\right]+\\
&+12 {R_0^B} {R_1^A} {R_1^B} T_1
\cos2 (\varPhi_1^A\hspace{-1mm}-\hspace{-1mm}\varphi_\varepsilon)  \cos2 (\varPhi_1^B\hspace{-1mm}-\hspace{-1mm}\varphi_\kappa ) \cos2(2 \varPhi_0^B\hspace{-1mm}-\hspace{-1mm}\varphi_\varepsilon \hspace{-1mm}-\hspace{-1mm}\varphi_\kappa )-\hspace{-1mm}\\
&\hspace{-1mm}-\hspace{-1mm}2 {R_1^D}^2 T_0 T_1 \cos^22 (\varPhi_1^D\hspace{-1mm}-\hspace{-1mm}\varphi_\kappa )\hspace{-1mm}-\hspace{-1mm}3 {R_0^B}^2T_1^2 \cos^22(2 \varPhi_0^B\hspace{-1mm}-\hspace{-1mm}\varphi_\varepsilon \hspace{-1mm}-\hspace{-1mm}\varphi_\kappa )+\\
&+4 {R_1^A}^2 {R_1^D}^2
\cos^22(\varPhi_1^A\hspace{-1mm}-\hspace{-1mm}\varphi_\varepsilon) \cos^22 (\varPhi_1^D\hspace{-1mm}-\hspace{-1mm}\varphi_\kappa )+\\
&+12 {R_1^B} {R_1^D} \cos2 (\varPhi_1^B\hspace{-1mm}-\hspace{-1mm}\varphi_\varepsilon) \cos2 (\varPhi_1^D\hspace{-1mm}-\hspace{-1mm}\varphi_\kappa ) \left[{R_0^B} T_1 \cos2(2 \varPhi_0^B\hspace{-1mm}-\hspace{-1mm}\varphi_\varepsilon \hspace{-1mm}-\hspace{-1mm}\varphi_\kappa )-\hspace{-1mm}\right.\\
&\left.\hspace{-1mm}-2 {R_1^A} {R_1^B} \cos2 (\varPhi_1^A\hspace{-1mm}-\hspace{-1mm}\varphi_\varepsilon)  \cos2 (\varPhi_1^B\hspace{-1mm}-\hspace{-1mm}\varphi_\kappa)
\right]>0\ \ \ \forall\fe,\fk;
\end{split}
\ee
this one can be further developed through extensive use of trigonometric formulae and introducing the angles $\Phi_A,\Phi_D,\delta_A$ and $\delta_D$ previously defined, along with the angle $\Phi_B=\Phi_0^B-\Phi_1^B$, an invariant of $\B$, and writing the condition for $\theta=\Phi_1^B$ instead than for $\theta=0$, like done until now (to do this, it is sufficient, in the polar formalism, to subtract the angle $\Phi_1^B$ from all the polar angles, see \cite{vannucci_libro}, p. 146). The result is
\be
\label{eq:conM4_1}
\begin{split}
&\left\{\left[T_0T_1+T_1R_0^A\cos4(\Phi_A-\delta_A-\fe)-2{R_1^A}^2\cos^22(\delta_A+\fe)\right]\times\right.\\
&\left.\times\left[T_0T_1+T_1R_0^D\cos4(\Phi_D-\delta_D-\fk)-2{R_1^D}^2\cos^22(\delta_D+\fk)\right]\right\}+\\
&+36{R_1^B}^4\cos^22\fe\cos^22\fk-6T_0T_1{R_1^B}^2(\cos^22\fe+\cos^22\fk)-\\
&-3T_1^2{R_0^B}^2\cos^22(2\Phi_B-\fe-\fk)-\\
&-24R_1^A{R_1^B}^2R_1^D\cos2(\delta_A+\fe)\cos2\fe\cos2(\delta_D+\fk)\cos2\fk-\\
&-6T_1{R_1^B}^2\left[R_0^A\cos4(\Phi_A-\delta_A-\fe)\cos^22\fk+\right.\\
&\left.+R_0^D\cos4(\Phi_D\hspace{-1mm}-\hspace{-1mm}\delta_D\hspace{-1mm}-\hspace{-1mm}\fk)\cos^22\fe\right]+12T_1R_0^BR_1^B\cos2(2\Phi_B\hspace{-1mm}-\hspace{-1mm}\fe\hspace{-1mm}-\hspace{-1mm}\fk)\times\\
&\times\hspace{-1mm}\left[R_1^A\cos2(\delta_A+\fe)\cos2\fk\hspace{-1mm}+\hspace{-1mm}R_1^D\cos2(\delta_D+\fk)\cos2\fe\right]>0\ \forall\fe,\fk.
\end{split}
\ee
This form is particularly interesting because the term in curly braces is actually the product of the condition $M2$ for $\A$ times the same condition for $\D$, cf. eqs.  (\ref{eq:condM2_1}) and  (\ref{eq:condM2_1bis}). The rest of the above expression depends on the invariants of $\B$, i.e. on $R_0^B,R_1^B$ and $\Phi_B$; this part represents hence the influence of coupling on the elastic  bounds of $\A$ and $\D$. Whenever $\B=\mathbb{O}$, all this part vanishes and condition $M4$ becomes redundant, as obvious, because it reduces simply to conditions (\ref{eq:condM2_2}) and (\ref{eq:condM2_3}), to be satisfied simultaneously. We remark also that, like for the case of $M3$, eq. (\ref{eq:conM4_1}) depends only on tensor invariants and on the two shift angles $\delta_A$ and $\delta_D$, so it is frame independent too.

Unfortunately, it is not possible to derive from the  condition (\ref{eq:conM4_1}) any explicit bound for the elastic moduli. In fact, on the one hand a procedure like the one described in the previous section for $M3$ is no longer possible in this case and, on the other hand, the problem could be tackled looking for the minimum of the expression in eq. (\ref{eq:conM4_1}), i.e. calculating the gradient with respect to $\fe$ and $\fk$, then solving the equation obtained equating it to zero, searching the minimum among the solutions and imposing such a minimum to be positive. Trying to do this leads to some trigonometric equations that cannot be solved analytically, though this remains possible numerically, for any application case. 

\section{Special cases}
Besides the conditions found above, the anisotropic moduli of $\A,\B$ and $\D$, i.e. $R_0^A,R_0^B,R_0^D,$ $R_1^A,R_1^B,R_1^D$ must be put non negative, as each one of them is the modulus of a complex quantity, \cite{vannucci_libro}, p. 144. Actually, $T_0$ and $T_1$ are also positive quantities, but no conditions are to be written for them because they are two elastic invariant moduli of the basic layer composing the laminate, so as moduli of a real material, they automatically satisfy the elastic bounds. 

All the previous conditions resume to the set of inequalities
\be
\label{eq:condgeneral}
\begin{split}
&R_0^A\geq0,\\
&R_1^A\geq0,\\
&R_0^B\geq0,\\
&R_1^B\geq0,\\
&R_0^D\geq0,\\
&R_1^D\geq0,\\
&T_0-R_0^A>0,\\
&T_1(T_0^2-{R_0^A}^2)-2{R_1^A}^2(T_0-R_0^A\cos4\Phi_A)>0,\\
&T_0T_1-{R_1^A}^2-3{R_1^B}^2>0,\\
%&(T_0T_1-{R_1^A}^2-3{R_1^B}^2)^2-(T_1R_0^A\sin4\Phi_A+3{R_1^B}^2\sin4\delta_A)^2-\\
%&-(T_1R_0^A\cos4\Phi_A-3{R_1^B}^2\cos4\delta_A-{R_1^A}^2)^2>0,\\
&T_1^2\left(T_0^2-{R_0^A}^2\right)+6T_1R_0^A{R_1^B}^2\cos4\left(\Phi_A+\delta_A\right)-\\
&-2{R_1^A}^2\left[T_1\left(T_0-R_0^A\cos4\Phi_A\right)-3{R_1^B}^2\right]-\\
&-6{R_1^B}^2\left(T_0T_1+{R_1^A}^2\cos4\delta_A\right)>0\\
&\min[M4]>0.
\end{split}
\ee 
In fact, eqs. (\ref{eq:condM2_3}) and (\ref{eq:condM3D}) are redundant for establishing the positive definiteness of matrix $[M]$. %, but not eqs. (\ref{eq:condM2_3})$_{1,2}$, needed to ensure the positiveness of the anisotropic part of $\D$. 
Though the last of the previous conditions cannot be given in an analytical explicit form, something more can be said in some particular cases, detailed below.

We notice also that the condition on $M1$ concerns just the material, and it is automatically satisfied, that on $M2$ concerns exclusively $\A$ (or alternatively $\D$), the one on $M3$ regards $\A$ (or $\D$) and $\B$ together and finally the condition on $M4$ concerns all the three tensors, $\A,\B$ and $\D$ together. Finally, it is immediate to check that whenever $\B=\mathbb{O}$ the conditions reduce simply to those already known, for $\A$ and for $\D$ separately:
\be
\label{eq:condgeneraldecoup}
\begin{split}
&R_0^A\geq0,\\
&R_1^A\geq0,\\
&T_0-R_0^A>0,\\
&T_1(T_0^2-{R_0^A}^2)-2{R_1^A}^2(T_0-R_0^A\cos4\Phi_A)>0,\\
&R_0^D\geq0,\\
&R_1^D\geq0,\\
&T_0-R_0^D>0,\\
&T_1(T_0^2-{R_0^D}^2)-2{R_1^D}^2(T_0-R_0^D\cos4\Phi_D)>0.
\end{split}
\ee 
This actually means that not only the matrix in eq. (\ref{eq:fundlaw}) or, which is equivalent, matrix $[M]$, eq. (\ref{eq:matrixM}), must be positive definite, but also $\A$ and $\D$ separately, as well known for uncoupled laminates, but not $\B$. The bounds involving moduli of $\B$ are not found imposing its positive definiteness:  $\B$ is actually not definite.

\subsection{Aligned orthotropic tensors}
\label{sec:aligort}
Let us consider the case of {\it aligned orthotropic tensors}, i.e. 
\be
\label{eq:alignedorth}
\begin{split}
&\delta_A=\lambda_A\frac{\pi}{2}, \ \delta_D=\lambda_D\frac{\pi}{2},\ \Phi_A=k_A\frac{\pi}{4},\ \Phi_b=k_B\frac{\pi}{4},\ \Phi_D=k_D\frac{\pi}{4},\\
& \lambda_A,\lambda_D,k_A,k_B,k_D\in\{0,1\}.
\end{split}
\ee
In such a case, very common in practice, it is easy to check that 
\be
\nabla M4=0\ \iff\ \fe=h_\eps\frac{\pi}{4},\ \fk=h_\kappa\frac{\pi}{4},\ h_\eps,h_\kappa\in\{0,1\},
\ee
i.e. the minimum of $M4$ can be only in one of the four points
\be
P_1=(0,0),\ P_2=\left(\dfrac{\pi}{4},0\right),\ P_3=\left(0,\dfrac{\pi}{4}\right),\ P_4=\left(\dfrac{\pi}{4},\dfrac{\pi}{4}\right).
\ee
Calling, for the sake of conciseness, $M4_1,M4_2,M4_3$ and $M4_4$ the value taken by (\ref{eq:conM4_1}) at $P_1,P_2,P_3$ and $P_4$ respectively, then the condition on $M4$ can be put in the form
\be
\min\{M4_1,M4_2,M4_3,M4_4\}>0,
\ee
where
\be
\label{eq:M4i}
\begin{split}
\hspace{-5mm}M4_1&=\left[T_0T_1+\hspace{-1mm}(-1)^{k_A}T_1R_0^A\hspace{-1mm}-\hspace{-1mm}2{R_1^A}^2\right]\left[T_0T_1+\hspace{-1mm}(-1)^{k_D}T_1R_0^D-2{R_1^D}^2\right]+\\
&+36{R_1^B}^4-3T_1^2{R_0^B}^2-6T_1{R_1^B}^2\left[2T_0+(-1)^{k_A}R_0^A+(-1)^{k_D}R_0^D\right]+\\
&+12(-1)^{k_B}T_1R_0^BR_1^B\left[(-1)^{\lambda_A}R_1^A+(-1)^{\lambda_D}R_1^D\right]-\\
&-24(-1)^{\lambda_A}(-1)^{\lambda_D}R_1^A{R_1^B}^2R_1^D,\\
\hspace{-5mm}M4_2&=T_1\left[T_0-(-1)^{k_A}R_0^A\right]\left[T_1\left(T_0+(-1)^{k_D}R_0^D\right)-2{R_1^D}^2-6{R_1^B}^2\right],\\
\hspace{-5mm}M4_3&=T_1\left[T_0-(-1)^{k_D}R_0^D\right]\left[T_1\left(T_0+(-1)^{k_A}R_0^A\right)-2{R_1^A}^2-6{R_1^B}^2\right],\\
\hspace{-5mm}M4_4&=T_1^2\left\{\left[T_0-(-1)^{k_A}R_0^A\right]\left[T_0-(-1)^{k_D}R_0^D\right]-3{R_0^B}^2\right\}.
\end{split}
\ee
In such a situation, the conditions (\ref{eq:condM2_2}) on $M2$ become
\be
\label{eq:condorth1}
\begin{split}
&T_0-R_0^A>0,\\
&\left[T_0-(-1)^{k_A}R_0^A\right]\left[T_1\left(T_0+(-1)^{k_A}R_0^A\right)-2{R_1^A}^2\right]>0,
\end{split}
\ee
while those on $M3$, eqs. (\ref{eq:condM3_3}) and (\ref{eq:condM3_3bis}),
\be
\label{eq:condorth2}
\begin{split}
&T_0T_1-{R_1^A}^2-3{R_1^B}^2>0,\\
&T_1\left[T_0-(-1)^{k_A}R_0^A\right]\left[T_1\left(T_0+(-1)^{k_A}R_0^A\right)-2{R_1^A}^2-6{R_1^B}^2\right]>0.
\end{split}
\ee
Because $T_1>0$ and by eqs. (\ref{eq:condorth1})$_1$ and (\ref{eq:condorth2})$_2$, condition (\ref{eq:condorth1})$_2$ is redundant and the above conditions can be rewritten as
\be
\label{eq:condM2M3}
\begin{split}
&T_0-R_0^A>0,\\
&T_0T_1-{R_1^A}^2-3{R_1^B}^2>0,\\
&T_1\left[T_0+(-1)^{k_A}R_0^A\right]-2{R_1^A}^2-6{R_1^B}^2>0.
\end{split}
\ee
A scrutiny of $M4_2$ and of $M4_3$ allows to see that if conditions (\ref{eq:condM2M3})$_{1,3}$ are considered in  eqs. (\ref{eq:M4i})$_{2,3}$, this will lead to two conditions that are exactly the analogous of those that could be obtained  by eqs.(\ref {eq:condM2_3}) and (\ref{eq:condM3D}), i.e. writing the conditions  $M2$ and $M3$ also for $\D$. 

Finally the whole set of conditions for the case at hand will be
\be
\label{eq:condorthaligned}
\begin{split}
&R_0^A\geq0,\\
&R_1^A\geq0,\\
&R_0^B\geq0,\\
&R_1^B\geq0,\\
&R_0^D\geq0,\\
&R_1^D\geq0,\\
&T_0-R_0^A>0,\\
&T_0T_1-{R_1^A}^2-3{R_1^B}^2>0,\\
&T_1\left[T_0+(-1)^{k_A}R_0^A\right]-2{R_1^A}^2-6{R_1^B}^2>0,\\
%&T_0-R_0^D>0,\\
%&T_0T_1-{R_1^D}^2-3{R_1^B}^2>0,\\
%&T_1\left[T_0+(-1)^{k_D}R_0^D\right]-2{R_1^D}^2-6{R_1^B}^2>0,\\
&\min\{M4_1,M4_2,M4_3,M4_4\}>0.
\end{split}
\ee
It is likely that  the above conditions constitute general bounds, i.e. that they are valid also for other more general situations, or, in other words, that the aligned orthotropic case is the worst one. However, this is just a conjecture that cannot be proved.

\subsection{Square symmetric $\B$}
Let us now consider the case of a square-symmetric coupling $\B$, i.e. a laminate designed to have $R_1^B=0$. In such a case, eq. (\ref{eq:condgeneral})$_9$ reduces to
\be
T_0T_1-{R_1^A}^2>0,
\ee
which is redundant, as previously discussed in Sect. \ref{sec:M2}. Moreover, eq. (\ref{eq:condgeneral})$_{10}$ becomes exactly eq. (\ref{eq:condgeneral})$_8$, so it can be discarded too. Actually, in this special case it is the whole set of conditions on $M3$ that it is redundant. This is quite natural, because when $R_1^B=0$ the only parameter accounting for coupling disappears from $M3$, which reduces by consequence to conditions already accounted for by $M2$. Moreover, eq. (\ref{eq:conM4_1}) reduces to
\be
\label{eq:conM4_R0}
\begin{split}
&\left[T_0T_1+T_1R_0^A\cos4(\Phi_A-\delta_A-\fe)-2{R_1^A}^2\cos^22(\delta_A+\fe)\right]\times\\
&\times\left[T_0T_1+T_1R_0^D\cos4(\Phi_D-\delta_D-\fk)-2{R_1^D}^2\cos^22(\delta_D+\fk)\right]-\\
&-3T_1^2{R_0^B}^2\cos^22(2\Phi_B-\fe-\fk)>0\ \forall\fe,\fk.
\end{split}
\ee
Also in this case it is interesting to consider the case of orthotropic aligned tensors, eq. (\ref{eq:alignedorth}), putting conventionally $\Phi_1^B=\Phi_0^B$ (if $R_1^B=0, \Phi_1^B$ is not defined). Repeating {\it verbatim} the steps already done in Sect. \ref{sec:aligort} gives eventually the conditions
\be
\label{eq:R1B0aligned}
\begin{split}
&R_0^A\geq0,\\
&R_1^A\geq0,\\
&R_0^B\geq0,\\
&R_0^D\geq0,\\
&R_1^D\geq0,\\
&T_0-R_0^A>0,\\
&T_1\left[T_0+(-1)^{k_A}R_0^A\right]-2{R_1^A}^2>0,\\
%&T_0-R_0^D>0,\\
%&T_1\left[T_0+(-1)^{k_D}R_0^D\right]-2{R_1^D}^2>0,\\
&\min\{M4_1,M4_2,M4_3,M4_4\}>0.
\end{split}
\ee
with now
\be
\label{eq:M4iR1B0}
\begin{split}
M4_1&=\left[T_1(T_0+(-1)^{k_A}R_0^A)-2{R_1^A}^2\right]\times\\
&\times\left[T_1(T_0+(-1)^{k_D}R_0^D)-2{R_1^D}^2\right]-3T_1^2{R_0^B}^2,\\
&\hspace{-8mm}M4_2=
T_1\left[T_0-(-1)^{k_A}R_0^A\right]\left[T_1(T_0+(-1)^{k_D}R_0^D)-2{R_1^D}^2\right],\\
&\hspace{-8mm}M4_3=
T_1\left[T_0-(-1)^{k_D}R_0^D\right]\left[T_1(T_0+(-1)^{k_A}R_0^A)-2{R_1^A}^2\right],\\
\hspace{-5mm}M4_4&=T_1^2\left\{\left[T_0-(-1)^{k_A}R_0^A\right]\left[T_0-(-1)^{k_D}R_0^D\right]-3{R_0^B}^2\right\}.
\end{split}
\ee

\subsection{Fully square symmetric laminates}
A more particular case, and even more interesting for applications, is that of {\it fully square symmetric laminates}, i.e. of 
\be
\label{eq:squsym}
R_1^A=R_1^B=R_1^D=0,
\ee
which is, e.g., automatically get if $R_1=0$, \cite{vincenti01}, i.e. if the basic layer is square symmetric itself, like actually it is for the very common case of layers reinforced by balanced fabrics. In such a case, the previous conditions (\ref{eq:R1B0aligned}) reduce to only
\be
\label{eq:R1aligned}
\begin{split}
&R_0^A\geq0,\\
&R_0^B\geq0,\\
&R_0^D\geq0,\\
&T_0-R_0^A>0,\\
&\min\{M4_1,MA_2,M4_3,M4_4\}>0.
\end{split}
\ee
with
\be
\label{eq:M4iR1B0}
\begin{split}
M4_1&=\left[T_0+\hspace{-1mm}(-1)^{k_A}R_0^A\right]\left[T_0+(-1)^{k_D}R_0^D\right]-3{R_0^B}^2,\\
&\hspace{-8mm}M4_2=
\left[T_0-(-1)^{k_A}R_0^A\right]\left[T_0+(-1)^{k_D}R_0^D\right],\\
&\hspace{-8mm}M4_3=
\left[T_0-(-1)^{k_D}R_0^D\right]\left[T_0+(-1)^{k_A}R_0^A\right],\\
\hspace{-5mm}M4_4&=\left[T_0-(-1)^{k_A}R_0^A\right]\left[T_0-(-1)^{k_D}R_0^D\right]-3{R_0^B}^2.
\end{split}
\ee
This special case of coupled laminates is interesting for applications because condition (\ref{eq:squsym}) is sufficient to obtain coupled thermally stable laminates, \cite{vannucci12joe1}, i.e. coupled laminates that preserve their form also under a temperature change, like the one occurred during the curing phase of pre-preg layers.

\subsection{$R_0$-orthotropic laminates}
Let us consider now the case of a laminate designed to have 
\be
R_0^A=R_0^B=R_0^D=0,
\ee
which is a special case of orthotropy, named $R_0$-{\it orthotropy}, \cite{vannucci02joe,vannucci10joe}. This kind of laminates can be obtained simply stacking layers with $R_0=0$. For this special case, eqs. (\ref{eq:condgeneral}) reduce to (we recall that like $T_1$, also $T_0>0$ automatically, because it is the modulus of a real material)
\be
\label{eq:condR0}
\begin{split}
&R_1^A\geq0,\\
&R_1^B\geq0,\\
&R_1^D\geq0,\\
&T_0T_1-2{R_1^A}^2>0,\\
&T_0T_1-{R_1^A}^2-3{R_1^B}^2>0,\\
&T_0^2T_1^2-2{R_1^A}^2(T_0T_1-3{R_1^B}^2)-6{R_1^B}^2(T_0T_1+{R_1^A}^2\cos4\delta_A)>0,\\
&\min\left\{\left[T_0T_1-2{R_1^A}^2\cos^22(\delta_A+\fe)\right]\left[T_0T_1-2{R_1^D}^2\cos^22(\delta_D+\fk)\right]+\right.\\
&+36{R_1^B}^4\cos^22\fe\cos^22\fk-6T_0T_1{R_1^B}^2(\cos^22\fe+\cos^22\fk)-\\
&\left.-24R_1^A{R_1^B}^2R_1^D\cos2(\delta_A+\fe)\cos2\fe\cos2(\delta_D+\fk)\cos2\fk]\right\}>0.
\end{split}
\ee 
If once again we consider the case of aligned tensors, eq. (\ref{eq:alignedorth}), then proceeding like in Sect. \ref{sec:aligort}, we get that condition (\ref{eq:condR0})$_4$ is redundant and finally
\be
\label{eq:condorthalignedR0}
\begin{split}
&R_1^A\geq0,\\
&R_1^B\geq0,\\
&R_1^D\geq0,\\
&T_0T_1-{R_1^A}^2-3{R_1^B}^2>0,\\
&T_0T_1-2{R_1^A}^2-6{R_1^B}^2>0,\\
&\min[M4_1,M4_2,M4_3]>0,
\end{split}
\ee
with now
\be
\label{eq:M4iR0}
\begin{split}
M4_1&=\left(T_0T_1-2{R_1^A}^2\right)\left(T_0T_1-2{R_1^D}^2\right)-\\
&-12{R_1^B}^2\left[T_0T_1-3{R_1^B}^2+2(-1)^{\lambda_A}(-1)^{\lambda_D}R_1^AR_1^D\right],\\
&\hspace{-8mm}M4_2=T_0T_1\left[T_0T_1-2{R_1^D}^2-6{R_1^B}^2\right],\\
&\hspace{-8mm}M4_3=T_0T_1\left[T_0T_1-2{R_1^A}^2-6{R_1^B}^2\right].
\end{split}
\ee
To notice that in this case $M4_4=T_0^2T_1^2>0$ always.

\subsection{Coupled isotropic laminates}
A last particular and rather interesting case is that of coupled laminates having an isotropic behavior in extension and in bending. Such a kind of laminates can be obtained in different ways, e.g. applying the Werren and Norris conditions, \cite{werren53} to laminates of the quasi-trivial type with $\A=\D$, \cite{vannucci01ijss,compsctech01}; an example of this kind of plates is the 18-layers laminate whose stacking sequence is
\begin{equation*} [0^\circ,  60^\circ,   -60_2^\circ,   60_2^\circ,    -60^\circ,   0^\circ,   60_2^\circ,      0^\circ,   -60^\circ,   0^\circ,   -60^\circ,   0_2^\circ,    -60^\circ,   60^\circ].\end{equation*}
In such a situation, it is $R_0^A=R_1^A=R_0^D=R_1^D=0$, while all the angles $\Phi_A,\Phi_D,\delta_A,\delta_D$ are no longer defined. To notice that it is necessarily $\A=\D$, because they are reduced to the only isotropic part, i.e. to $T_0$ and $T_1$, that are those of the basic layer, while $\B$ is necessarily anisotropic, because $T_0^B=T_1^B=0,R_0^B+R_1^B\neq0.$

In this situation, eq. (\ref{eq:conM4_1}) becomes
\be
\label{eq:conM4iso}
\begin{split}
&T_0^2T_1^2+36{R_1^B}^4\cos^22\fe\cos^22\fk-6T_0T_1{R_1^B}^2(\cos^22\fe+\cos^22\fk)-\\
&-3T_1^2{R_0^B}^2\cos^22(2\Phi_B-\fe-\fk)>0\ \forall\fe,\fk,
\end{split}
\ee
eq. (\ref{eq:condgeneral})$_9$ is redundant with respect  to eq. (\ref{eq:condgeneral})$_{10}$ and finally conditions (\ref{eq:condgeneral}) reduce to only
\be
\label{eq:coniso}
\begin{split}
&R_0^B\geq0,\\
&R_1^B\geq0,\\
&T_0T_1-6{R_1^B}^2>0,\\
&\min[M4]>0.
\end{split}
\ee
Once more, if $\B$ is orthotropic, i.e. $\Phi_B=k_B\dfrac{\pi}{4},k_B\in\{0,1\}$, then
\be
\begin{split}
&M4_1=T_1^2(T_0^2-3{R_0^B}^2)-12{R_1^B}^2(T_0T_1-3{R_1^B}^2),\\
&M4_2=M4_3=T_0T_1(T_0T_1-6{R_1^B}^2),\\
&M4_4=T_1^2(T_0^2-3{R_0^B}^2),
\end{split}
\ee
and conditions (\ref{eq:coniso}) become simply
\be
\begin{split}
&R_0^B\geq0,\\
&R_1^B\geq0,\\
&T_0T_1-6{R_1^B}^2>0,\\
&\min\{M4_1,M4_2,M4_4\}>0.
\end{split}
\ee

\section{Conclusion}
Coming back to the questions posed in Sect. \ref{sec:intro}, the result found above show that:
\begin{enumerate}[i.]
\item it is possible to establish some bounds involving {\it also} the moduli of $\B$ but not {\it exclusively} these ones;
\item the moduli of $\A$ and $\D$ known for the case of uncoupled laminates are still valid but in addition some other bounds, relating the moduli of $\A,\B$ and $\D$ are added in the case of coupled laminates;
\item not all the bounds relating the moduli of  $\A,\B$ and $\D$ can be given in an explicit form;
\item all the bounds depend exclusively on tensor invariants and shift angles, so all of them are frame independent;
\item the existence of some kind of symmetries affects the bounds and normally allows to express them in a simpler form. 
\end{enumerate}
The previous results show also that  $\B$ remains an undefined tensor. 
%Finally, the presence of coupling induces some additional requirements, involving at the same time the moduli of $\A,\B$ and $\D$; in other words, the presence of coupling imposes some interactions with the moduli of $\A$ and $\D$, that are in this way affected by supplementary conditions to be satisfied. 
The bounds found in this paper reveal hence the existence of some supplementary conditions to be satisfied by the moduli of $\A$ and $\D$, these conditions imposing some sort of interaction with the moduli of $\B$. This result, unknown until now, is interesting {\it per se}, because it shows that coupling is not unbounded, but also in practical applications, namely in optimization problems, to correctly define the feasibility domain in the search of optimal coupled laminates. 
%There are, as often happens,  other ways to approach the same complicated problem, and some of them have been explored by the author, but all of them necessarily lead to a greater number of conditions.

%%
  \bibliographystyle{vancouver} 
 \bibliography{biblio}

\end{document}